\documentclass[11pt]{amsart}
\usepackage[foot]{amsaddr}

\usepackage[a4paper, margin=3cm]{geometry}
\usepackage{graphicx}
\usepackage{amssymb}
\usepackage{epstopdf}
\usepackage{csvsimple}
\usepackage{float}
\usepackage[labelfont=bf, font=small]{caption}

\title{Infections and Identified Cases of COVID-19 from Random Testing Data}
\author{Allen Caldwell, Vasyl Hafych, Oliver SChulz, Lolian Shtembari \\ Max Planck Institute for Physics, Munich }

\date{May 15, 2020}                 

\begin{document}

\begin{abstract} There are many hard-to-reconcile numbers circulating concerning Covid-19.  Using reports from random testing, the  fatality ratio per infection is evaluated and used to extract further information on the actual fraction of infections and the success of their identification for different countries.
\end{abstract}

\maketitle

\section{Introduction}
There are many hard-to-reconcile numbers circulating concerning the new corona virus.  A wide range  for the fraction of the population with positive tests for Covid-19 has been reported, with orders of magnitude differences in the case fatality ratio (CFR) in different countries. Getting a better understanding of the fraction of the population that has been infected with Covid-19 and of its lethality is of utmost importance for guiding further actions.  This paper presents an analysis of publicly available random testing data and is aimed at providing information on the number of infections and the lethality of the novel corona virus.  This is quantified as an infected fatality ratio (IFR). We perform this study as an exercise in data analysis, without attempting to interpret or modify reported numbers.  The data that we use is the data we found up until May 12, 2020. \\
% The lethality of Covid-19 is very important to understand as a guide to the type of confinement measures that should be used:  clearly, a high lethality would indicate strong measures are necessary and makes controlling the epidemic even more important.  

One use of the analysis carried out in this note is to evaluate the probability that infected  individuals were identified. This probability is important to gauge the effectiveness of proposed contact tracing measures to control the spread of the virus.  We use the extracted IFR results from the random testing data to evaluate this probability in a large sample of countries and find a wide variation in the reported fraction of infected individuals. \\

The data from random testing that is available in some form to-date (in several cases as reports in newspaper articles) are employed to address these various questions.  This is not a fine-grained analysis - only the extraction of probable ranges for our quantities of interest for large populations is attempted.  It is also quite preliminary, as there is limited random testing data available and the level at which this data is truly representative of the broader population is not clear. Nevertheless, we find that the reported data are largely self-consistent and therefore offer some guidance.

\section{The Data}
The data considered for the extraction of the IFR of Covid-19 is the available data as of May 12, 2020 from random tests for antibodies.  We assume that the presence of antibodies in previously infected people is ascertainable with high probability three weeks after infection (typical numbers reported are about 85~\%).  We then take for our estimate of the IFR the number of deaths approximately one week after the reported dates of the random testing for antibodies.   The choice of using one week is based on the estimate that the time lag between developing severe symptoms and death is on average approximately 12 days (see e.g.~\cite{ref:LosAlamos}) and the assumption that the development of significant antibodies occurs on average some time after the development of the symptoms. Our results will depend on the correctness of this assumption, and we estimate how our results could have changed if we had picked a different time lag in the discussion on systematic uncertainties in section~\ref{sec:Systematics}. \\

  Random testing has also been performed to detect the presence of the virus (see ~\cite{ref:Austria,ref:Iceland,ref:Sweden,ref:Kansas}), but these data are more difficult to use in extracting the IFR.  The reason is that the detection of the virus is limited to a time window when the virus is present in the location being tested (throat, lungs) and that only the deaths related to this time period should then be used to evaluate the IFR.  We leave this analysis to a later study. \\

Seven random testing samples for antibodies were found, three of which have been well documented~\footnote{We have recently become aware of a study undertaken in Spain that would result in a much higher IFR value than the values we find.  At the moment, we do not have the information needed but the results can be added to our study once a more complete sets of numbers are reported.}. These data sets are given in Table~\ref{default} and are the basis for this analysis. It is intended that this analysis will be updated as improved data becomes available.

\begin{table}[!h]
\caption{Data used in the analysis (see text for references).  The symbol $N$ is used to represent the number of people tested for antibodies, while $k$ is the number of positive results. The numbers on population are from readily accessible web resources.  The numbers of deaths are from either from the reports summarizing the studies or from web data made available by the local authorities. }
\begin{center}
\begin{tabular}{|r|r|r|r|r|r|c|}
\hline\hline
\textbf{Study} & $\textbf{N}$ & $\textbf{k}$   & \textbf{Population} & \textbf{Deaths} \\
\hline
Santa Clara & 3330 
 & 50 
 & 1928000 
 & 50
 \\
Chelsea & 200 
 & 64 
 & 40160 
 & 39 
 \\
Kreis Heinsberg & 919 
 & 142 
 & 12597 
 & 7 
 \\
LA County & 863 
 & 35 
 & 10040000
 & 600
 \\
New York  & 7500 
 & 1118 
 & 19450000 
 & 17500
\\
Miami & 1800 
 & 108 
 & 2717000 
 & 287 
 \\
Geneva & 576
 & 50
 &  499480
 & 243 
 \\

\hline
\end{tabular}
\end{center}
\label{default}
\end{table}%

\subsection{Data treatment}
The data were taken at face value where possible.  In several cases, further analysis of the reported numbers were discussed in the publications leading to revisions of the raw numbers.  We do not apply these corrections here but note the size of the effect for reference and discuss these in the section on systematic uncertainties later in the paper.  In some cases, estimates were necessary since not all required information was available.  The following should be taken into account:\\

\begin{itemize}
\item The raw number of positive tests in Santa Clara was 50 out of 3330~\cite{ref:SantaClara}, which is about 1.5~\%.  The authors of the study corrected the observed fraction to account for locality, sex and ethnicity and found a result of 2.8~\%; i.e., a factor of two higher than the raw observed value.  This would lead to a lower IFR if this value was used.  This correction was not taken into account in the central values extracted for this analysis but were considered as a systematic variation.  The tests were performed on April 3,4.  The number of deaths one week later was 50, while 83 deaths had been recorded two weeks after performing the tests.  It is relevant to note that the infected fraction, being quite low, is very sensitive to the specificity of the test, which can introduce a large systematic uncertainty. \\
\item  The dates at which the tests were performed in Chelsea, Massachusetts~\cite{ref:Chelsea} could not be found but are assumed to be in the first week of April. The tests excluded patients who had tested positive for the virus from nasal swabs.  Of the 200 people tested, 64 tested positive for antibodies.  This large fraction of positive tests relaxes the uncertainty due to the specificity.  The number of deaths was taken on April 20th.  This could possibly be too late, resulting in an overestimate of the IFR.\\
\item The Kreis-Heinsberg results were taken from the latest values reported in \cite{ref:Heinsberg}.  The `effective' number of positive tests was taken as 15.5~\% of the number of individuals tested, as this is the fraction of positive tests reported in the reference.  We note that this value is the result of a detailed analysis by the authors.   The number of deaths in Gangelt, the town in which the tests were performed, was 7 at the time of the tests and increased by 1 in a two week follow up period.  Using this larger number of deaths would result in a 15~\% increase in the IFR value from this particular study.  However, since the numbers are quite small this  change is already accounted for in the large uncertainties associated with our extracted IFR (see next section).\\
\item The number of positive tests in Los Angeles county were estimated based on the reported fraction of positive tests in \cite{ref:LosAngeles}, where 4.1~\% of the tested cases were reported to have the antibody to the virus. We estimated the number of positive tests based on this value. The tests were performed on April 10,11. This study is purported to represent all of Los Angeles county, which had 550 reported deaths on April 15th and 797 ten days later.  For our analysis, we took the number of deaths to be 600, corresponding to the approximate value one week after the tests.\\
\item Two different sets of numbers have been reported for New York - one for New York City and one for New York State.  Here the statewide numbers are taken. The number of positive tests in New York State was estimated based on the reported fraction of positive tests in \cite{ref:NewYork}.  The testing was started on April 20 and the fraction of positive tests for the antibodies was reported as `nearly 14.9~\%' from 7500 tests.  The number chosen for the analysis was 1118 positive cases.  The number of deaths on April 27th was approximately 17500.  The number of deaths around this date was increasing by about 350 per day, or 2~\% per day.  This data is the most statistically important in the study.  As we will see, it also results in the highest IFR value of the 6 data sets.\\
\item The number of positive tests in Miami (Dade County) was estimated based on the reported fraction of positive tests in \cite{ref:Miami}.  There, it is said that `nearly 1800 individuals have participated in the program' and `6~\%' tested positive for antibodies.  The tests were carried out in the time window from April 2-8, and the number of deaths reported as due to Covid-19 is taken from April 18th - i.e., it is likely an overestimate of the IFR for the Miami area. \\
\item In studies conducted in Geneva, random samples of the population were tested on consecutive weeks and early results of the study have been made available~\cite{ref:Geneva}.  We use the results for the testing performed in the third week (the week of April 20, 2020), where 576 individuals were tested, resulting in 50 positive tests.  As of April 30, 243 deaths had been reported for a population of nearly 500,000.  The specificity of the antibody tests was very high, with no false positives in a sample of 176 known negative cases, and a true positive rate of 86.2~\%. The authors used these values to evaluate a somewhat higher infected fraction than seen in the raw numbers (9.7~\% versus 8.7~\%).  As in the other cases, we do not correct for this difference.
\end{itemize}

%The values in Table~\ref{default} can be displayed in different ways.  One choice is to plot the ratio of the number of positive tests for antibodies to the number of cases tested and this is displayed in Fig.~\ref{fig:rand_sampl_data} for the available sets of tests.  
%As can be seen in the figure, there is a wide range in the fraction of positive tests, from 1.5~\% in Santa Clara to approximately 30~\% in the Chelsea study.  The numbers of tested individuals are typically small, so that appropriate statistical treatment is necessary.  Furthermore, although the tests are initially intended to represent random samplings of the population, there are surely biases.  The authors are not competent to discuss  possible biases.

%\begin{figure}[!t]
%    \centering
%    \includegraphics[width =0.9\textwidth]{fig-1.pdf}
%    \caption{The fraction of positive tests from the randomized %tests available to date (left) and the observed case fatality ratio (right).}
%    \label{fig:rand_sampl_data}
%\end{figure}

\clearpage
\section{Analysis - IFR}
\label{sec:IFR}

The basic assumption made in this article is that, for a large population base and for populations with similar levels of medical care, the IFR from Covid-19 should not differ widely once the population has been infected across all classes of individuals evenly (e.g., across age groups).  There are clearly very significant differences in IFR amongst age groups, and many other factors play crucial roles in individual cases such as medical preconditions or the availability of top-rate medical services.  It is assumed that these will tend to average out for large (say 100,000 or more) population groups in many countries so that it is meaningful to talk about the average lethality of the disease for these areas.  This averaging has likely not yet occured in many areas as we are still in the relatively early days of the pandemic, so that our results will not be truly representative in individual cases.  However, it should be the case that by taking a number of different studies these effects tend to average out.  More discussion on the biases that can result can be found in section~\ref{sec:Systematics}. \\

Given our basic assumption, it becomes possible to track the average number of infected people based on the number of recorded deaths due to Covid-19.  To arrive at the population averaged IFR results, the data in Table~\ref{default} were analyzed as follows.  First, the probability of a positive test in a random sample was taken to follow a Binomial distribution:

\begin{equation}
\label{eq:Binomial}
P(k | N,f) = \binom{N}{k} f^k  (1-f)^{N-k} 
\end{equation}

\noindent where $f$ represents the probability that a person in the area of interest was infected, $N$ is the number of random tests performed, and $k$ is the number of positive tests.  The value of $f$ clearly varies widely across the different areas as seen in Table~\ref{default}.  We note that the value of $k$ also depends on the efficiency of the test in positive cases, and can also be overestimated by false positive cases (specificity of the tests).  We do not attempt to correct for these  effects and assume here that they do not significantly bias the results.  Expert knowledge in these issues would be needed for this, which we do not possess.  We discuss the possible variations in the results in Section~\ref{sec:Systematics}. \\

The probability of having a number of deaths $D$ given a population size $P$ is then taken to follow a Poisson probability distribution as the IFR is  known to be a small number.  We have that the expected number of infected people in the sample is:

$$I = f \cdot P $$

\noindent and the expected number of deaths due to Covid-19 from this number of infected people will be (after the appropriate lag time):

$$\lambda = r \cdot I$$

\noindent where here $r$ is the IFR value for the particular sample.  We note  again that the number of deaths is counted significantly later than the number of infected people (typically 7 days later in our analysis).  The probability of observing $D$ deaths is then:

$$P(D | \lambda) = \frac{e^{-\lambda} \lambda^D}{D!}  $$

In terms of the parameters that we use in the analysis, we write this as:

\begin{equation}
\label{eq:Poisson}
P(D | f,r,P) = \frac{e^{-rfP} (rfP)^D}{D!} 
\end{equation}
\\
The number of reported deaths due to Covid-19 is subject to different definitions and should also be considered as uncertain.  Again, we do not attempt a correction for these variations as we do not have the competence for this. Possible variations in our results due to this uncertainty in Section~\ref{sec:Systematics}.\\

\subsection{Probability distribution for the IFR}

To implement the  condition that the IFR should not differ too broadly in different areas, the individual IFR values, $r_i$ in the different regions were assumed to come from a `parent distribution' that is meant to represent the range of possible IFR results. This parent distribution should clearly have zero probability for an IFR value of zero, and cannot be larger than a few ~\% based on known results.  We therefore choose distributions that can represent this behavior.  Our primary results are derived using a log-normal distribution, and the Weibull distribution is used as an alternative to test the systematic uncertainties pertaining to our choice. \\

  It is important to realize that {\bf the distribution that we extract for the IFR is the main result of the analysis}.  I.e., we are not expecting a single value for the IFR in all regions, but expect that all values are within roughly a factor 2-3 of each other, with variations depending on the particular conditions in the area under consideration.  The extracted distribution should represent in some approximate way the distribution of IFR across a wide range of conditions.  Once we extract the parameters of this distribution, we then use it to define a range of infections for different countries in the next section.
\\

The log-normal probability distribution that is used to describe the IFR probability in different regions is defined as:

\begin{equation}
\label{eq:Lognormal}
P(r) = \frac{1}{r \sigma \sqrt{2\pi}} \exp \left[ - \frac{1}{2} \left(\frac {\ln (r/\mu) +\sigma^2/2}{\sigma}\right)^2\right] \;\;\; .
\end{equation}

\begin{figure}[!t]
    \begin{center}
    \includegraphics[width =0.75 \textwidth]{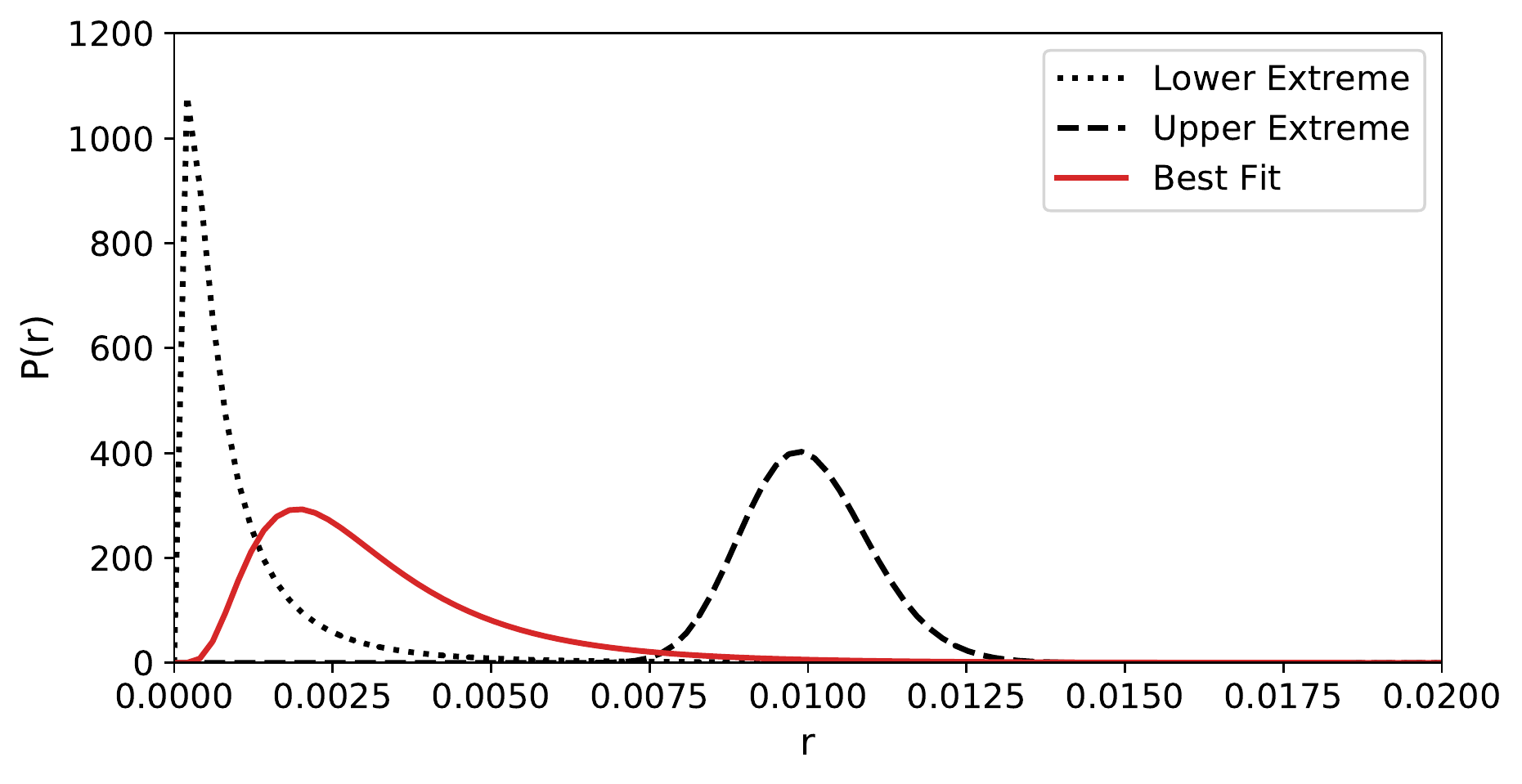}
    \end{center}
    \caption{IFR distribution from the random testing data.  The log-normal distributions for the extremes of the parameter ranges allowed in the analysis are shown in the dotted and dashed curves. All log-normal distributions with shape between these extremes are allowed in the analysis.  The shape of the distribution from the fit to the data in Table~\ref{default} using the most probable values of the probability distributions for $\mu$ and $\sigma$ is shown as the solid curve.}
\label{fig:lognormal}
\end{figure}

This distribution introduces two parameters in the analysis, $\mu$ and $\sigma$, that control the log-normal distribution.   The parameter $\mu$ is the mean of the log-normal distribution while $\sigma$ controls the width of the distribution. The possible shapes of the log-normal distribution considered in our analysis are shown in Fig.~\ref{fig:lognormal}.
\\

The analysis is carried out using Bayes Theorem to yield the probability distributions on the parameters of interest:

$$
P(\mathbf{r,f},\mu,\sigma| \mathbf{k,N,P,D}) = \frac{P(\mathbf{D,k}|\mathbf{r,f},\mathbf{N,P})  
P(\mathbf{r,f},\mu,\sigma)}
{\int P(\mathbf{D,k}|\mathbf{r,f},\mathbf{N,P})  
P(\mathbf{r,f},\mu,\sigma) d(\mathbf{r,f},\mu,\sigma)} 
$$
\noindent where the bold-faced symbols represent vectors of numbers.  The analysis was carried out using the Bayesian Analysis Toolkit (see \cite{ref:BAT}).  
The individual terms in our expression are given as follows: we have that
$$P(\mathbf{D,k}|\mathbf{r,f},\mathbf{N,P}) = P(\mathbf{k}|\mathbf{f},\mathbf{N}) P(\mathbf{D}|\mathbf{f,r},\mathbf{P}) $$
where the two terms on the right hand side of the equation are products of terms of the kind given in Equations~\ref{eq:Binomial} and ~\ref{eq:Poisson}. 
The second term in the numerator is:
$$
P(\mathbf{r,f},\mu,\sigma)=P_0(\mathbf{f})P(\mathbf{r}|\mu,\sigma)P_0(\mu)P_0(\sigma) \; .
$$
The prior probabilities (those with subscript $0$) are all taken as flat prior probabilities and $P(\mathbf{r}|\mu,\sigma)$ is given by the product of expressions of the form given in Eq.~\ref{eq:Lognormal}.
\\

\begin{figure}[!t]
    \begin{center}
    \includegraphics[width =0.9 \textwidth]{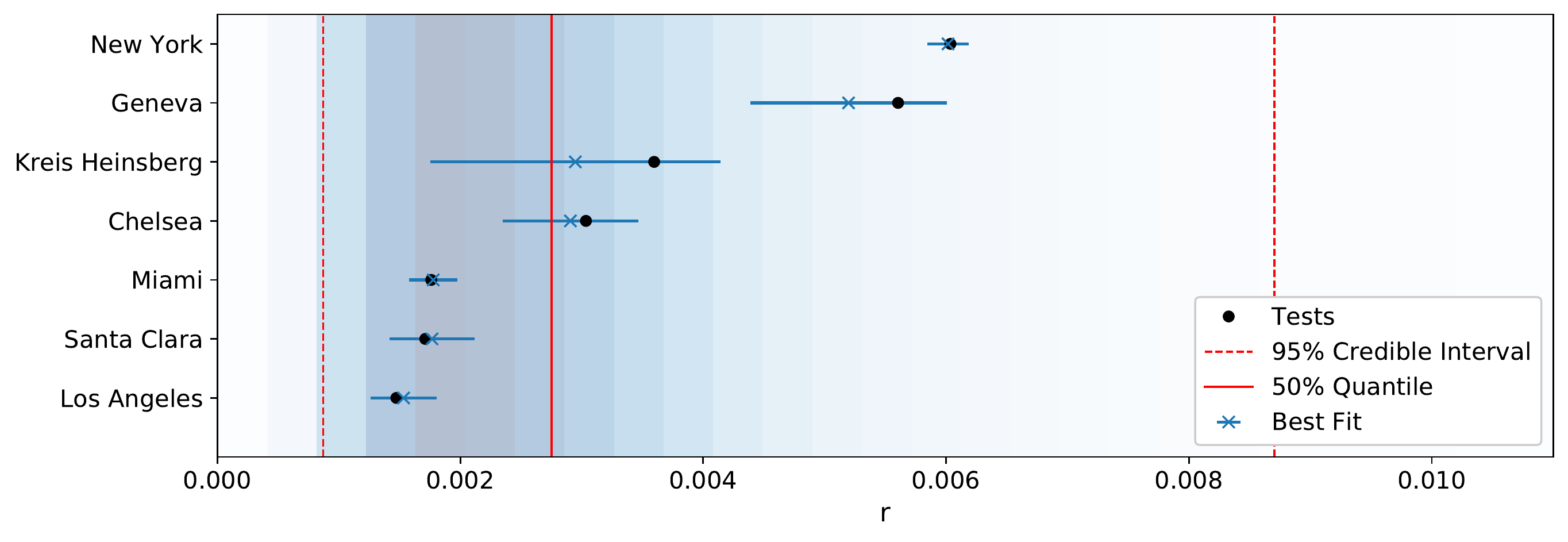}
    \end{center}
    \caption{The results of the infected fatality ratio (IFR) analysis.  The black symbols represent the values obtained from a simple scaling of the results in Table~\ref{default}.  The crossed symbols show the IFR values evaluated for the individual area studies (i.e., the individual $r_i$ values) together with the associated 95~\% central probability interval for $r_i$. The result for the overall lFR distribution is also shown as the shaded background in the figure.  The dashed red vertical lines show the 95~\% central probability interval while the median value of IFR from the log-normal distribution is shown as the solid line.}
\label{fig:lognormal_and_rand_smpl}
\end{figure}

Fitting the data in Table~\ref{default} yields the results shown in Fig.~\ref{fig:lognormal} and~\ref{fig:lognormal_and_rand_smpl}.  The shape of the log-normal distribution using the best-fit parameter values is shown in Fig.~\ref{fig:lognormal}. The best-fit distribution is comfortably between the limits set in the analysis.  As is seen in the plot, the peak of the distribution is for an IFR around 0.25~\%, with a tail out to approximately 0.75~\%.  The IFR values extracted from the random testing data and the total number of deaths recorded one week after the tests therefore lead to the conclusion that the IFR is generally below 1~\%.
\\

The results from the IFR analysis from the individual studies are displayed in Fig.~\ref{fig:lognormal_and_rand_smpl}.  The overall range of values of IFR from the log-normal distribution is shown together with the individual $r_i$ results extracted from the fit. For the individual fits, best fit values are shown together with the 95~\% uncertainties as horizontal bands.  The median and central 95~\% probability range of the IFR from the log-normal distribution are also shown in the plot, and the shaded background indicates the probability density of the IFR distribution. While there is some spread in the results derived from different studies, the range is not so large as that seen in the CFR values, with individual IFR values ranging from approximately 0.18~\% to 0.6~\%. 
\\

In addition to these results, the simple scaling result (deaths resulting from Covid-19) divided by (observed positive test fraction multiplied by the population of the area under study) are also shown.  
\\

 As a next step, we use these results to evaluate the infected fractions in different countries and also evaluate the effectiveness of diagnosing positive cases using currently used methods.

\clearpage
\section{Analysis - infections and diagnosis}
\label{sec:infected}

We now take the results from the IFR analysis and apply it to different countries to evaluate both the degree of infections as well as the effectiveness of recognizing infected and infectious cases.  For this, we use the data given in Table~\ref{table-2}.  The data were taken from \cite{ref:owid} and the population numbers are from \cite{ref:Population}.\\

%\begin{table}[!b]
%\captionof{table}{Data used in infection tracing and infected fraction analysis as of May 15, 2020.  The countries and their ISO codes are given, together with populations numbers from \cite{ref:Population}.  The symbol $I_{rec}$ gives the number of recorded infections, while $D$ is the number of recorded deaths due to Covid-19.}
%\begin{center}
%\begin{tabular}{|c|c|c|c|c|}\hline%
%Country & ISO & Population & $I_{rec}$ & $D$ \\\hline\hline
%\csvreader[late after line=\\\hline]%
%{table-2.csv}{CountryArea=\CountryArea, ISO=\ISO, %Population=\Population, RC=\RC, RD=\RD}%
%{ \CountryArea & \ISO & \Population & \RC & \RD }%
%\end{tabular}
%\end{center}
%\label{table-2}
%\end{table}

\begin{table}[!b]
\captionof{table}{Data used in infection tracing and infected fraction analysis as of May 15, 2020.  The countries and their ISO codes are given, together with populations numbers from \cite{ref:Population}.  The symbol $I_{rec}$ gives the number of recorded infections, while $D$ is the number of recorded deaths due to Covid-19.}
\begin{center}
\begin{tabular}{|r|r|r|r|r|}
\hline\hline
\textbf{Country} & \textbf{ISO} & \textbf{Population} & \textbf{$\textbf{I}_{\textbf{rec}}$} & \textbf{D} \\\hline
United States & USA & 331002000 & 1417889 & 85906 \\
United Kingdom & GBR & 67886911 & 233151 & 33614 \\
Italy & ITA & 60461826 & 223096 & 31368 \\
France & FRA & 65273511 & 141356 & 27425 \\
Spain & ESP & 46754778 & 230253 & 27321 \\
Belgium & BEL & 11589000 & 54288 & 8903 \\
Germany & DEU & 83783942 & 173152 & 7824 \\
Netherlands & NLD & 17134872 & 43481 & 5590 \\
Canada & CAN & 37742000 & 73401 & 5472 \\
China & CHN & 1439323776 & 84029 & 4637 \\
Turkey & TUR & 84207771 & 144749 & 4007 \\
Sweden & SWE & 10099265 & 28582 & 3529 \\
Switzerland & CHE & 8654622 & 30380 & 1588 \\
Romania & ROU & 19254711 & 16247 & 1046 \\
Poland & POL & 37846611 & 17615 & 883 \\
Japan & JPN & 126476461 & 16193 & 710 \\
Austria & AUT & 9006000 & 16005 & 626 \\
Denmark & DNK & 5792000 & 10713 & 543 \\
Hungary & HUN & 9663457 & 3417 & 442 \\
Finland & FIN & 5540720 & 6145 & 287 \\\hline\hline
\end{tabular}

\end{center}
\label{table-2}
\end{table}

\subsection{Estimating the delay between infection reports and death reports}
\begin{figure}[!t]
\begin{center}
\includegraphics[width =0.9\textwidth]{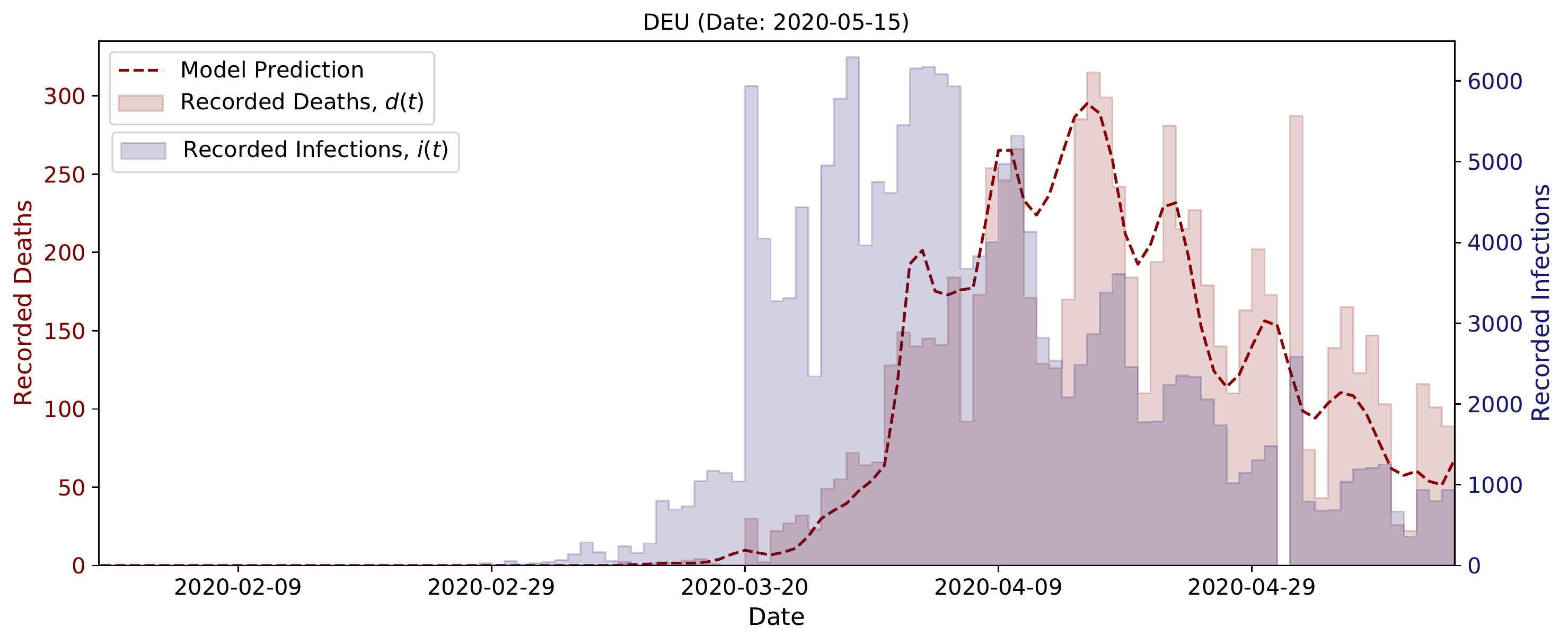}
\end{center}
\caption{The number of recorded infections (blue distribution) as a function of date for Germany.  The number of recorded deaths due to Covid-19 is shown as the red distribution.  The shifted, smeared and scaled prediction from the infected number is shown as the red dashed curve.}
\label{fig:delay}
\end{figure}

As a first step, we extract the relationship between the number of recorded deaths due to Covid-19 and the number of recorded infections.  The number of recorded deaths follow a similar development as the number of recorded infections, but is clearly delayed in time.  This can be seen in Fig.~\ref{fig:delay} where we show the results for Germany as an example.  The number of deaths can be predicted from the number of infections by forecasting the number of deaths as a fraction of the number of infected cases and distributing these over a number of days.  Mathematically, we predict the number of deaths on day T as:

\begin{equation}
\label{eq:future-death}
%d(t) = \sum_{\tau=0}^T S\cdot i(\tau) \cdot \mathcal{P}(t-(\tau+\Delta),\sigma_D)
\hat{d}(T) = \sum_{t=1}^T S\cdot i(t) \cdot \mathcal{P}(T|t+\Delta,\sigma_D)
\end{equation}

\noindent where $i(t)$ is the number of recorded infections on day $t$, $S$ is a scale factor to predict the number of deaths (the CFR), $\Delta$ quantifies a shift between the recorded infections and the recorded deaths, and $\sigma_D$ describes how the deaths are distributed and is given in days.  Note that we use lower case letters ($d,i$) to represent the daily death and recorded infection values, while the upper case letters are reserved for the integrated quantities.  I.e., we have that
\begin{eqnarray*}
D(T) &=& \sum_{t=1}^T d(t) \\
I(T) &=& \sum_{t=1}^T i(t) 
\end{eqnarray*}
The symbol $\hat{d}(T)$ is our estimate for the number of deaths on day $T$ according to our model.\\

A truncated normal distribution is used for $\mathcal{P}(T|t+\Delta,\sigma_D)$ to model the spread of the infected cases forward in time. The expression evaluated at date $T$ is
$$
\mathcal{P}(T|t+\Delta,\sigma_D)= \frac{\int_{T - 1}^{T} e^{-\frac{1}{2} \left(\frac{t'-t - \Delta}{\sigma_D}\right)^2}dt'}{\int_{t-1}^{\infty} e^{-\frac{1}{2} \left(\frac{t'-t-\Delta}{\sigma_D}\right)^2}dt'}
$$

In order to extract the relevant parameter values, a $\chi^2$ minimization is performed.  The result for Germany is shown in Fig.~\ref{fig:delay} and led to the parameters: $S=0.048$, $\Delta=13.6$~days and $\sigma_D=0.9$~days.  Recall that $S$ is the CFR, so we have a case fatality ratio of about 5~\% for Germany.\\

\begin{figure}[!t]
\begin{center}
\includegraphics[width =0.9 \textwidth]{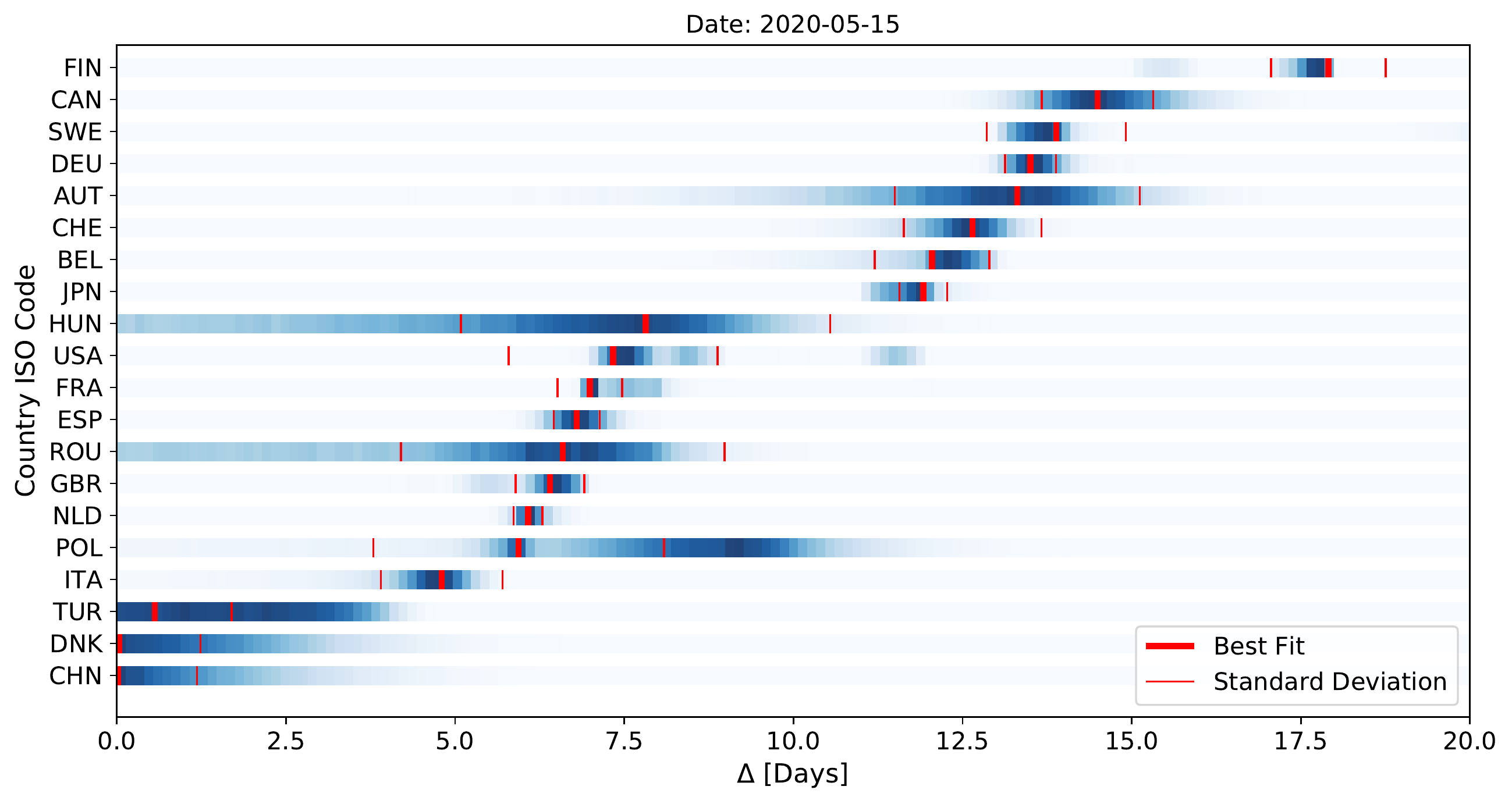}
\end{center}
\caption{The fitted delay, $\Delta$, between recorded number of infections and the recorded number of deaths due to Covid-19 for the different countries studied.  The most probable value for $\Delta$ is shown as the thick red line and the thin red lines indicate the central 95~\% probability intervals.  The blue shading indicates the probability density of the $\Delta$ distribution.}
\label{fig:alldelay}
\end{figure}

 The results for the parameter $\Delta$ are shown in Fig.~\ref{fig:alldelay} for all countries given in Table~2.  The figure shows the extracted probability density for $\Delta$ as the blue shaded band, and also the most probable values as well as one standard deviation intervals.  As is seen, the results can be divided roughly into three categories:  those countries with very small delay, a set of countries with delay of $5-7$ days (these are countries where typically a large number of infections were reported) and countries such as Germany where the delay is closer to 2 weeks.  The latter set of countries typically did not have a surge of infections that stressed the medical resources of the country.  We use these values of $\Delta$ to extract the date at which we will report the fraction of the infected population in the following.\\
 
\begin{figure}[!ht]
\begin{center}
\includegraphics[width =0.9 \textwidth]{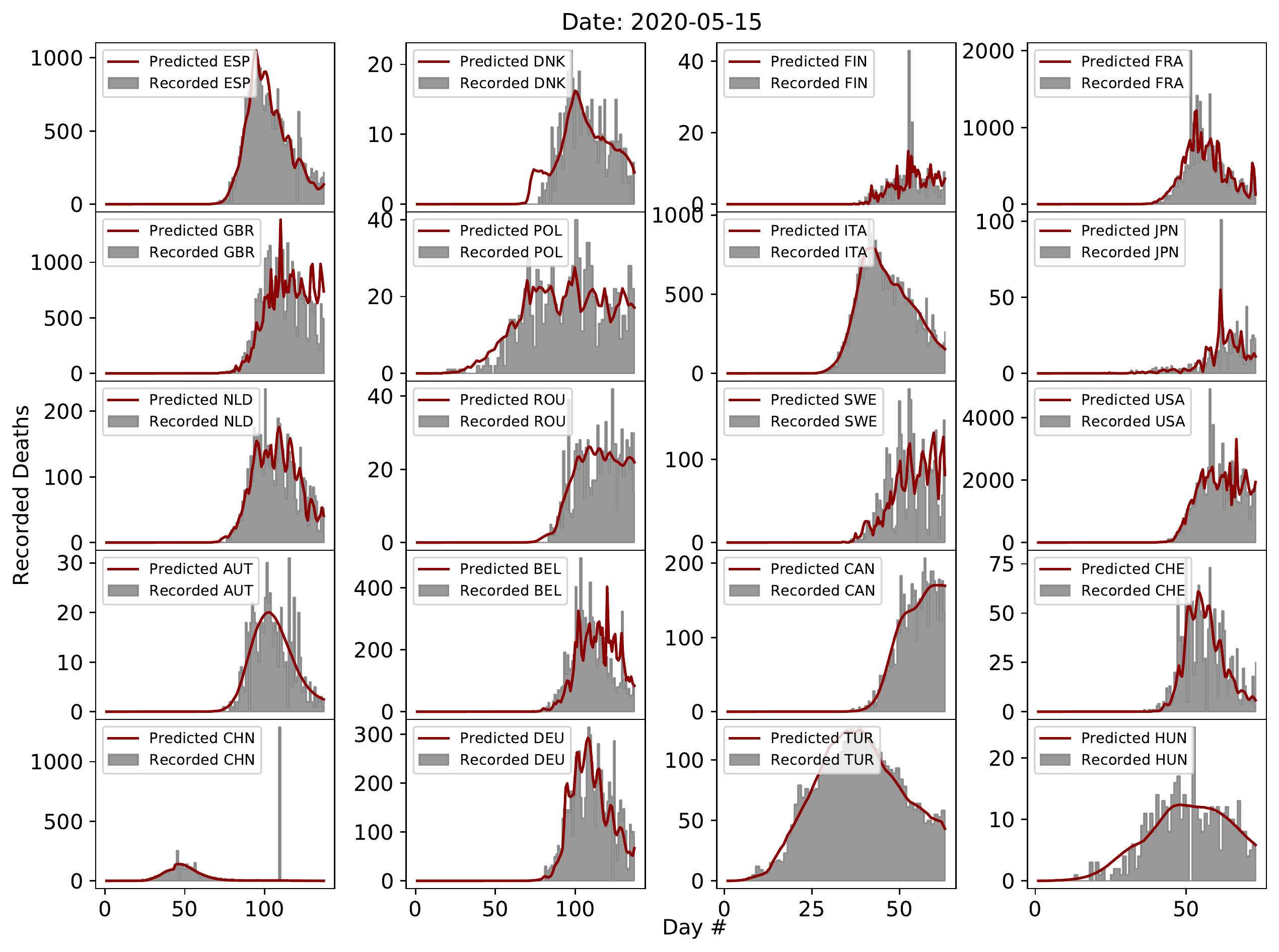}
\end{center}
\caption{The distributions for the number of deaths from the reported infections for all countries studied are displayed as histograms. The best fit distributions for the predictions based on the number of reported infections are shown as the red curves. The functional form used to predict the number of deaths is given in Eq.~\ref{eq:future-death}.}
\label{fig:all}
\end{figure}

The comparison of the predicted values of deaths using the expression in Eq.~\ref{eq:future-death} and the observed values are shown in Fig.~\ref{fig:all} for all countries in Table~2.  As can be seen in the figure, a reasonable description is achieved in most cases.  Some countries show spikes in the reported death distributions due to changes in counting procedures.  These spikes cannot be modeled by our simple equation and are not real effects but due to re-evaluation of standards for the counting.

\subsection{Estimating the fraction of the population infected with Covid-19}

The results from the delay analysis allow us to estimate the fraction of the population that has been infected at a fixed date by using the number of reported deaths and scaling up using the IFR range from the log-normal distribution:

\begin{equation}
\label{eq:past-total-deaths}
\hat{I}(T-\Delta) = \frac{D(T)}{r}
\end{equation}

\noindent where $\hat{I}(T-\Delta)$ is the estimated number of infected people at the date $(T-\Delta)$ and $D(T)$ is the number of reported deaths at date $T$.  The value of the IFR, $r$, follows the log-normal distribution whose parameters are extracted from the most likely values obtained from the fit of $\mu$ and $\sigma$ we performed in the previous sections~\footnote{We do not allow the estimated number of infected individuals to exceed the population of the country.}. Since the IFR value is taken from a probability distribution, we obtain a probability distribution for $\hat{I}$, from which we can extract the $95$~\% central interval limits.\\

The results of the analysis for our selection of countries are reported in Table~\ref{table-3}. The median infected fractions range from nearly $0$ for China to nearly $30$~\% for Belgium.  The  $95$~\% ranges span a factor of $10$.  I.e., there is a large uncertainty in the results coming from the fact that the IFR distribution is quite broad.  For most countries, we can conclude however that the percentage of the population that has been infected is in the single digits.\\

The limits that we extracted for the infected population of each country were based on the most recent number of deaths. As we have discussed earlier, the number of deaths is related to the number of infected individuals, but it also contains a time delay, quantified by $\Delta$. This means that if we extrapolate the infected population from the current number of deaths, we will obtain an estimate that is representative of the past number of infected people, specifically $\Delta$ days in the past.\\

It is possible to estimate $\hat{I}$ at the current date by using the number of deaths we expect in the future, $D(T + \Delta)$. We can estimate the cumulative distribution of $D(T + \Delta)$ by propagating forward in time the number of deaths using the fitted model we described above and the current number of reported individuals.  Substituting these new estimates in Equation~\ref{eq:past-total-deaths} we obtain:

\begin{equation}
\label{eq:present-total-deaths}
\hat{I}(T) = \frac{D(T + \Delta)}{r}
\end{equation}

\begin{figure}[!ht]
\begin{center}
\includegraphics[width =0.8 \textwidth]{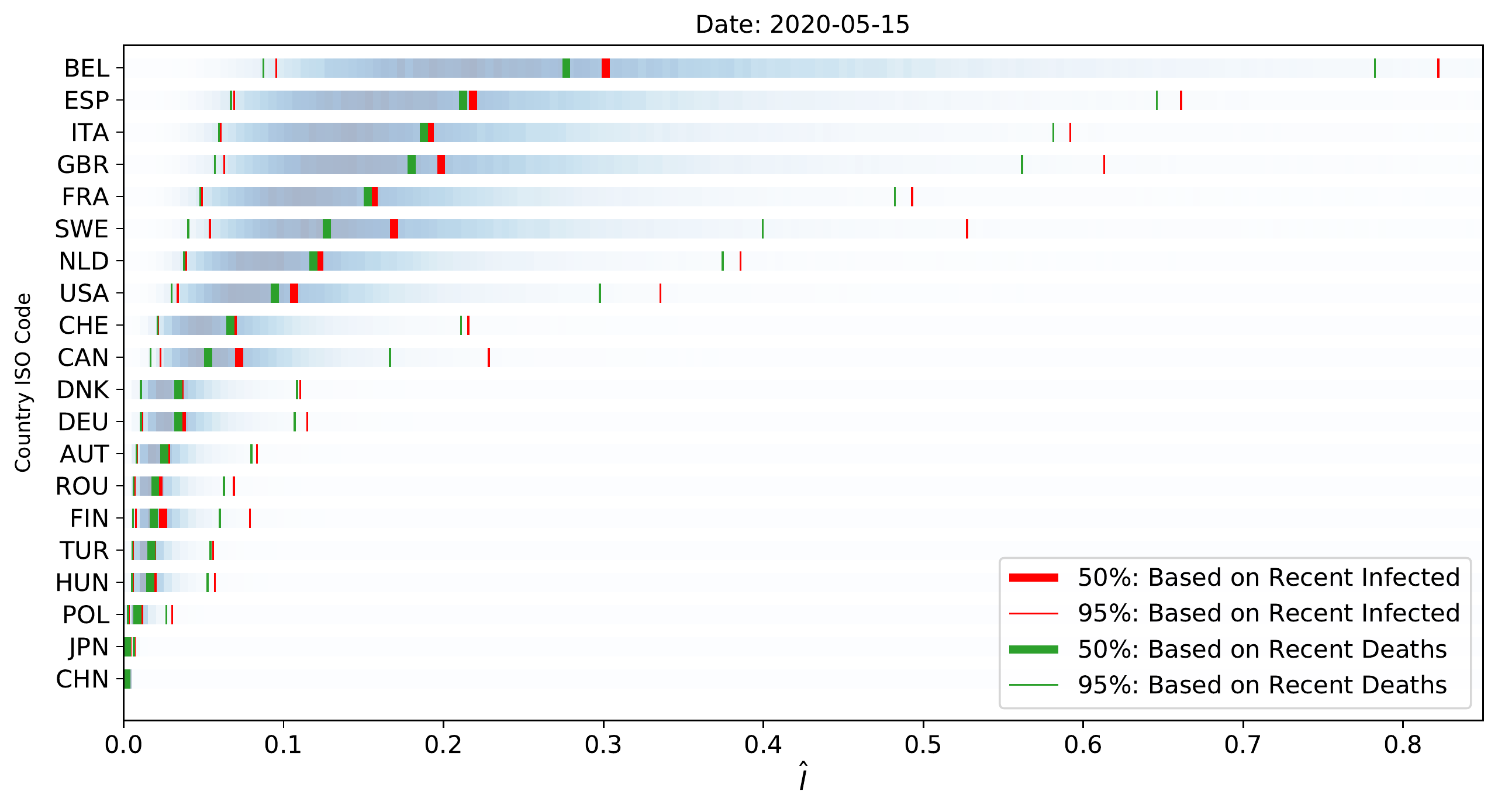}
\end{center}
\caption{The fractions of the population for the countries listed in Table~2 that have been infected with COVID-19.  The blue shaded bands indicate the probability density as a function of the estimated infected fraction. The solid green line indicates the median estimate at the date (May 15, 2020 - $\Delta$) where the delay is country dependent and is shown in Fig.~5. The thin green lines indicate the 95~\% central probability intervals.  The solid red lines indicate the estimated infected fraction on May 15, 2020 by propagating the number of expected deaths forward as described in the text.}
\label{fig:infected}
\end{figure}

Both estimates of $\hat{I}(T-\Delta)$ and $\hat{I}(T)$ are shown in Fig.~\ref{fig:infected}. 

%\begin{table}[!b]
%\captionof{table}{The range of estimated infected fraction and reported fractions of the population based on the IFR and $\Delta$ estimates. The values reported in the table correspond to the dates May 15, 2020 - $\Delta$ for the appropriate countries.  The columns show the 95~\% central probability intervals for the infected fraction of the population, $\hat{I}$ and the fraction of the infected population that has been reported as infected, $\hat{f}_R$.}
%\begin{center}
%\begin{tabular}{|c|c|c|}\hline%
%Country & $\hat{I}$, 95\% &  $\hat{f}_R$, 95\% 
%\\\hline\hline
%\csvreader[late after line=\\\hline]%
%{table-3.csv}{CountryArea=\CountryArea, FI=\FI, FR=\FR}%
%{ \CountryArea & \FI &  \FR }%
%\end{tabular}
%\end{center}
%\label{table-3}
%\end{table}

\begin{table}[!b]
\captionof{table}{The range of estimated infected fraction and reported fractions of the population based on the IFR and $\Delta$ estimates. The values reported in the table correspond to the dates May 15, 2020 - $\Delta$ for the appropriate countries.  The columns show the 95~\% central probability intervals for the infected fraction of the population, $\hat{I}$ and the fraction of the infected population that has been reported as infected, $\hat{f}_R$.}
\begin{center}
\begin{tabular}{|r|r|r|}
\hline\hline
\textbf{Country} & \textbf{$\hat{I}$, 95\%} & \textbf{$\hat{f}_R$, 95\%} \\\hline
USA & 0.034 - 0.34 & 0.012 - 0.13 \\
GBR & 0.063 - 0.61 & 0.0055 - 0.056 \\
ITA & 0.061 - 0.59 & 0.0061 - 0.06 \\
FRA & 0.049 - 0.49 & 0.0044 - 0.044 \\
ESP & 0.069 - 0.66 & 0.0071 - 0.071 \\
BEL & 0.095 - 0.82 & 0.0049 - 0.049 \\
DEU & 0.012 - 0.11 & 0.018 - 0.18 \\
NLD & 0.039 - 0.39 & 0.0066 - 0.066 \\
CAN & 0.023 - 0.23 & 0.0085 - 0.086 \\
CHN & 0.00037 - 0.0037 & 0.022 - 0.22 \\
TUR & 0.0056 - 0.056 & 0.03 - 0.3 \\
SWE & 0.054 - 0.53 & 0.0053 - 0.054 \\
CHE & 0.022 - 0.22 & 0.017 - 0.17 \\
ROU & 0.0069 - 0.069 & 0.012 - 0.12 \\
POL & 0.003 - 0.03 & 0.014 - 0.14 \\
JPN & 0.00069 - 0.0069 & 0.019 - 0.19 \\
AUT & 0.0084 - 0.083 & 0.021 - 0.21 \\
DNK & 0.011 - 0.11 & 0.016 - 0.16 \\
HUN & 0.0057 - 0.057 & 0.0059 - 0.06 \\
FIN & 0.0077 - 0.079 & 0.014 - 0.15 \\\hline\hline
\end{tabular}

\end{center}
\label{table-3}
\end{table}

\subsection{Reported fraction}
The fraction of the infected population that has been identified is then calculated as the ratio of positively diagnosed cases to the estimated total number of infections:
$$
\hat{f}_R(T-\Delta)=\frac{I_{\rm rec}(T-\Delta)}{\hat{I}(T-\Delta)}= \frac{I_{\rm rec}(T-\Delta)}{D(T))/r}=\frac{r\cdot I_{\rm rec}(T-\Delta)}{D(T)}
$$

\noindent where $I_{\rm rec}(T-\Delta)$ is the number of reported cases at the date $(T-\Delta)$, $D(T)$ is the reported number of deaths on the date $T$ and $r$ is the IFR value from the log-normal distribution.  In extracting our estimate for the reported fraction of infections, $\hat{f}_R(T_\Delta)$, we allowed the value of $\Delta$ and $r$ to vary according to the probability distributions from the fit yielding $\Delta$ and from the log-normal distribution, respectively.  This leads to a rather broad distribution for $\hat{f}_R(T_\Delta)$.\\

\begin{figure}[!hb]
\begin{center}
\includegraphics[width =0.8 \textwidth]{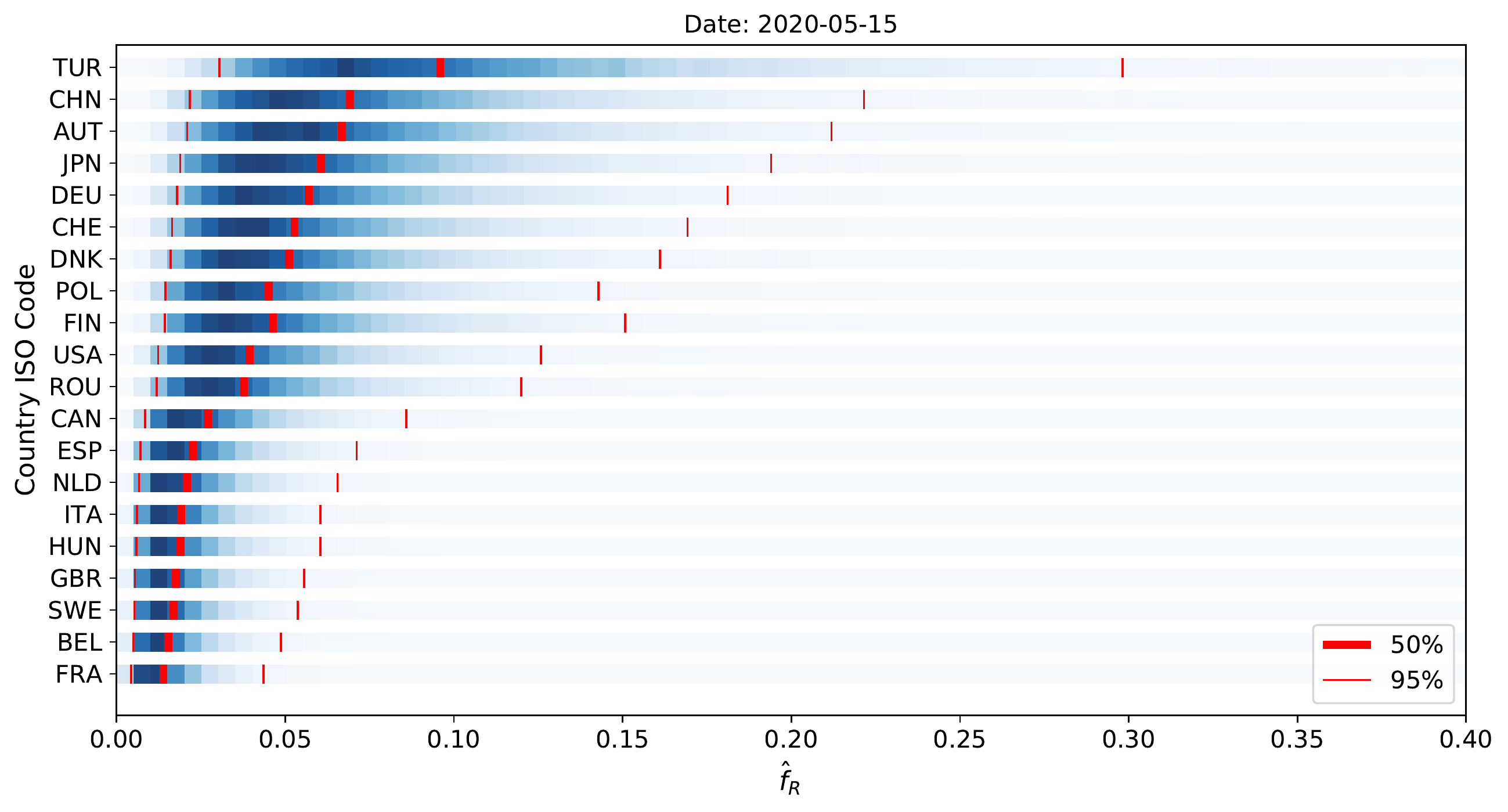}
\end{center}
\caption{The fractions of the infected population for the countries listed that have been positively identified and reported. The blue shaded bands indicate the probability density as a function of the estimated reporting fraction. The solid red line indicates the median estimate at the date (May 15, 2020 - $\Delta$) where the delay is country dependent and is shown in Fig.~5. The thin red lines indicate the 95~\% central probability intervals.}
\label{fig:reported}
\end{figure}

 The results of the analysis are summarized in Table~\ref{table-3} and shown graphically in Fig.~\ref{fig:reported}. The reported percentages of diagnosed cases are typically also estimated to be in the single digit range.  Note that the same IFR range was used for all countries.  If a given country is believed to have a lower IFR than the average, then for that country the reporting fraction will tend to be smaller, and vice-versa.

\clearpage

\section{Systematics}
\label{sec:Systematics}
A number of assumptions were made at various points in this analysis, and we estimate how our results vary when we take different assumptions into account.\\

\subsection{False positives in antibody tests}
As noted in Section~\ref{sec:IFR}, the number of reported infections in the random testing data is modified by false positive results and also by inefficiency of the tests to recognize genuine infections.  We have that, for a single test, the probability to have a positive result is

$$
P(+) = P(+|I)f + P(+| \bar{I})(1-f)
$$
where $P(+)$ is the probability of a positive test result, $P(+|I)$ is the probability of a positive test result given that a person was infected, $P(+| \bar{I})$ is the false  positive test result and $f$ is the probability that a person has been infected. Turning the expression around, we have

$$
f = \frac{P(+)- P(+| \bar{I})}{P(+|I)-P(+|\bar{I})} \; .
$$

The denominator is taken to be close to 1, so that

$$
f \approx P(+)- P(+| \bar{I}) \; .
$$

If the specificity of the test is high, meaning $P(+| \bar{I})$ is very small, and $P(+|I)\approx 1$, then we can assume that $f\approx P(+) $. 
For the Santa Clara study, we have that $P(+) \approx 0.015$.  A $0.5$~\% false positive rate would lead to a correction of $f$ downwards by approximately this value, or $0.5~\%$, which would increase the IFR value for Santa Clara by $50$~\%.  A false positive rate of $1$~\% would triple the IFR value, bringing it close to the New York result. At this point, we can also recall that the authors of the Santa Clara study argued that the correct rate is closer to 2.8~\% than 1.5~\% due to biases in the random sampling population, so the two corrections would go in opposite directions. \\

For the other studies, the values of the observed positive testing fraction is high enough that false positive rates below about $1$~\% will not significantly affect the results.\\

\subsection{Other reliability concerns of the random testing}
In order to extract an IFR that should represent an average over the full population, it is not sufficient that the random sampling represents the different population subgroups.  Indeed, it is also important that the distribution of those infected in different population categories (age group, health conditions, ...) is known since there are strong correlations with these aspects of the population. If, e.g., the fraction of those infected in the upper age brackets is disproportionately high in a region related to one of the studies, then the extracted IFR from a random sampling of that population will also be high.  This type of biasing must also be taken into account in extracting the final results.  Some of the studies mentioned (see e.g. ~\cite{ref:Heinsberg, ref:Geneva}) attempt to account for this type of biasing.  We have not done so here, so that our results could potentially be biased due to this type of effect.  Mathematically, we can represent the situation as follows:
$$
E[D] = \int r(a,h,o) f(a,h,o) p(a,h,o) d(a,h,o)
$$
where $E[D]$ is the expected number of deaths, $p(a,h,o)$ is the population density as a function of age ($a$), of health status, modeled here as a continuous quantity ($h$) and of possibility other relevant characteristics ($o$). This population density has associated an infected fraction $f(a,h,o)$ that can depend on all of these characteristics.  Our assumption in the paper is that the random sampling data follows this product of $f\cdot p $ approximately correctly so that we can estimate
$$
E[r(a,h,o)] \approx \frac{D_{\rm recorded}}{f\cdot P}
$$

The IFR result that we use in the analysis is rather broad and allows for factors of 2-3 differences and hopefully cover the remaining sampling uncertainties.\\

\subsection{How deaths are counted}
There are clearly differences in how deaths due to Covid-19 are counted in different regions.  Even within a single region, the definition can change as is clearly seen in Fig.~6, where spikes in the death count appear.  These uncertainties should be propagated in our analysis.  One mitigating factor is that we are using integrated values, such that short term fluctuations are averaged out.  However, revised death counts are typically in the upward direction, so an upward shift of the median IFR value (currently below 0.3~\% from the random testing data)  is likely.  For the extraction of our probability intervals, we have used the log-normal distribution for the IFR, which has a broad range and a tail to larger values.  We believe that this adequately accounts for this expected upward correction in the number of recorded deaths due to Covid-19.

\subsection{Time delay between tests and recorded deaths}
In this analysis, we have used a time delay between the date of the tests for antibodies and the date at which the number of deaths assigned to Covid-19 were counted of one week.  This delay was assumed based on modeling data~\cite{ref:LosAlamos} that indicates on average 12 days between hospitalization and death.  It is not clear at which point antibodies will show up in significant amounts in the course of the infection, and the one week estimate in counting the development of antibodies and the number of deaths could be flawed.   Taking the USA as an example, a one day change in the date at which deaths are counted would change the results by approximately 2.5~\%.  For cases where the infection was developing quickly, the number could be considerably higher.  We therefore assume that changes in the number of deaths could certainly be wrong by 25~\%.  However, the range that we assign to the IFR from the log-normal distribution is much broader than this, so that this uncertainty is not expected to significantly affect the results.\\

\subsection{Choice of the log-normal distribution}
\begin{figure}[!t]
    \begin{center}
    \includegraphics[width =0.65 \textwidth]{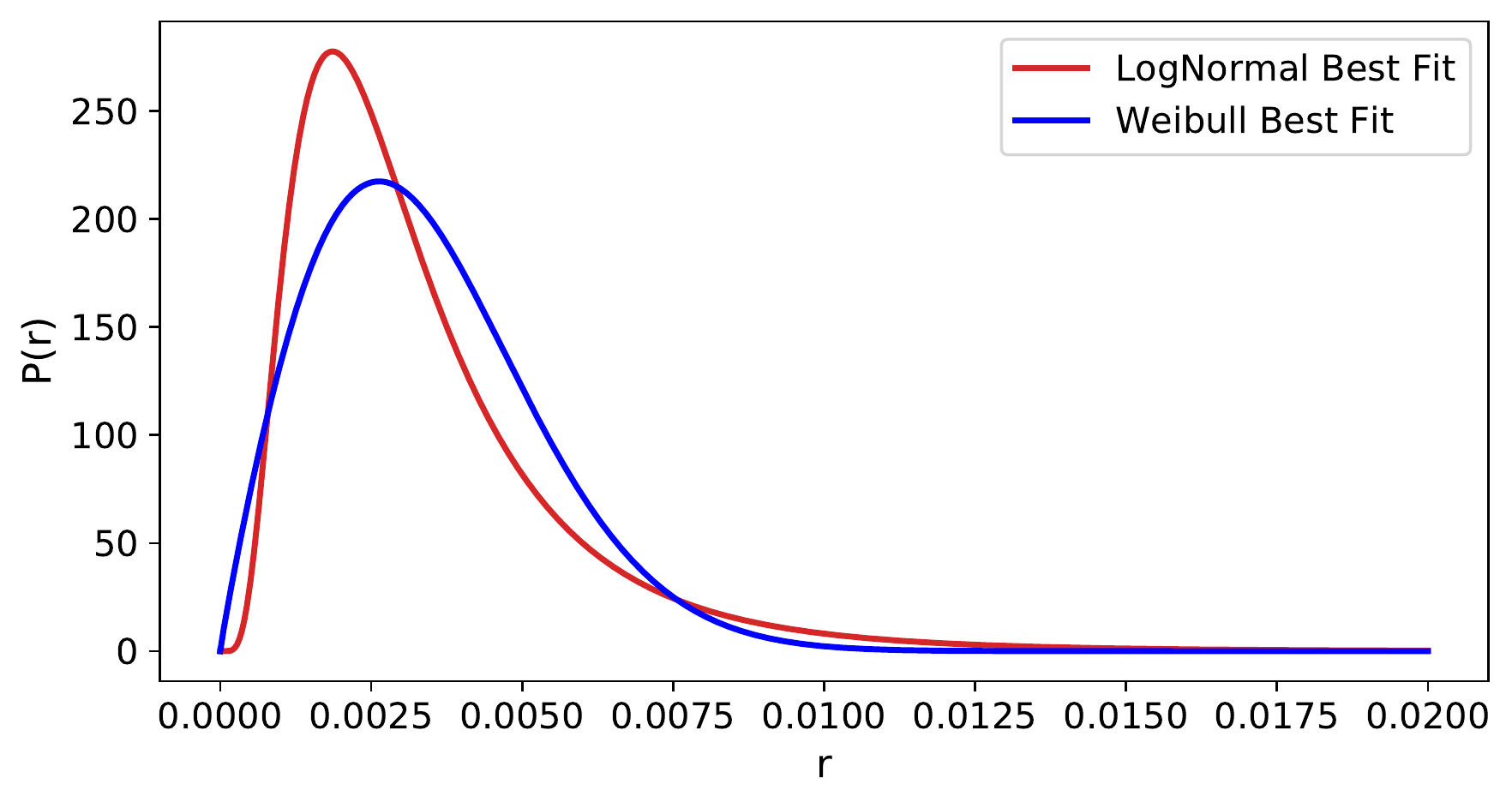}
    \end{center}
    \caption{The shape of the distributions from the fit to the data in Table~\ref{default} using the most probable values of the parameters for $(\mu, \sigma)$ and $(\lambda, k)$ are shown. In red the fit for the log-normal distribution  and in blue the fit for the Weibull distribution.}
\label{fig:lognormal-weibull}
\end{figure}

The log-normal distribution was chosen as the `parent' distribution due to its well-motivated shape.  Other functional forms can also have similar shapes, one of which is the Weibull distribution:

$$
P(r|\lambda,k) = \frac{k}{\lambda} \left(  \frac{r}{\lambda} \right)^{k-1} e^{-(x/\lambda)^k} \;\;\; x\geq 0 \;\; .
$$
Requiring $k>1$ leads to a distribution starting at $0$ for $x=0$ and with a similar shape as the log-normal distribution.  Extraction the range of IFR values from this distribution led to minimal changes in the results, as can be seen in Fig.~\ref{fig:lognormal-weibull}.\\

\subsection{Constancy of $\Delta$}

\begin{figure}[h]
\begin{center}
\includegraphics[width =0.7 \textwidth]{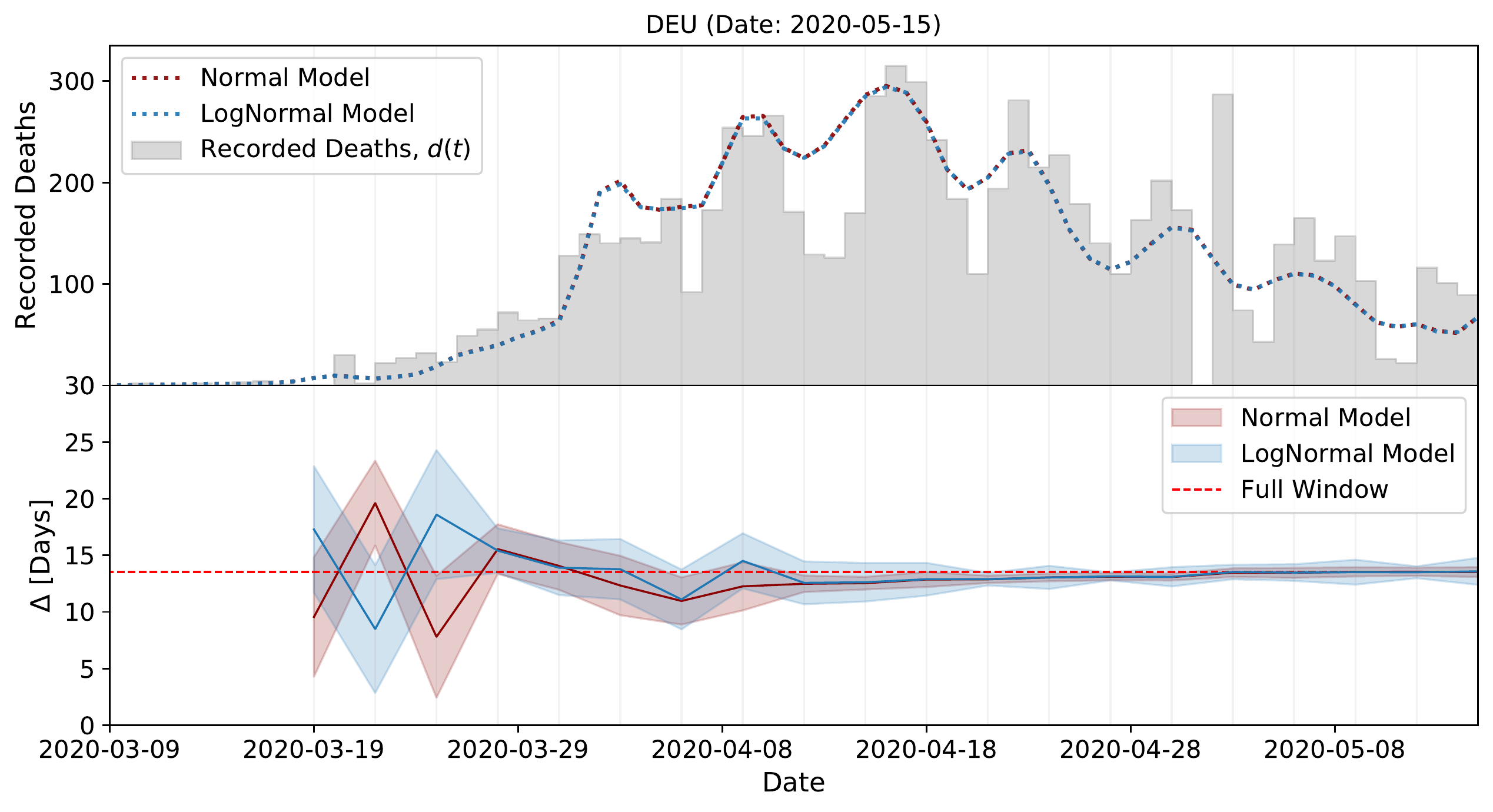}
\end{center}
\caption{Top: Comparison of the extraction of $\Delta$ for Germany using a log-normal versus a normal distribution.  The two different predictions are overlayed in the plot. Bottom: The result for $\Delta$ as the maximum of the time range is varied.   The shaded bands indicate the one standard deviation uncertainty on $\Delta$.}
\label{fig:stability}
\end{figure}

In order to test the stability of the predicted number of deaths due to different choices of their distribution in the future, we performed our analysis adopting a log-normal distribution instead of a truncated normal in Equations~\ref{eq:future-death}.
In the case of Germany, the results of this comparison are presented in Fig.~\ref{fig:stability} (top). We notice that the two prediction curves are practically indistinguishable from one another. Additionally, in order to test the stability of the estimated delay between infection reports and death reports, we investigate evolution of the parameter $\Delta$ by performing an analysis by limiting the time range to a specific day $T$. The results of this study are shown in Fig.~\ref{fig:stability} (bottom). As we increase $T$, we notice that our estimate for the mean of $\Delta$ approaches a constant value and the error of this measure, estimated as the standard deviation of $\Delta$, is also reduced in time.\\

\clearpage
\section{Conclusions}

An analysis of available random testing data has been performed in order to extract information on the lethality of the Covid-19 virus.  Assuming this lethality does not vary widely across large population groups, the information was used to extract the fraction of infected individuals in a number of countries and to also estimate the fraction of infected individuals that have been reported as positively identified.  A number of caveats apply:\\
\begin{itemize}
\item The random test samples used in many the IFR data sets are quite small \\
\item The data sets, although quoted as random, are likely biased samples of the population groups targeted and may not represent the distribution of the infected population \\
\item The results from this paper rely on the time lag between infection (reporting) and death.  The assumptions used may not be accurate.
\end{itemize}

In spite of these caveats, we find that there is good consistency in the extracted values for the infected fatality ratio, IFR, across the different random data sets.  We use the extracted IFR to estimate the fraction of the population that has been infected, and the fraction of those infected that have been positively identified.  We find a broad range of results for a sample of $20$ countries.  For the infected fraction, the typical percentages are in the single digits.  The reporting fraction tends to be closer to 5~\%. \\

The results presented in this simple analysis contain both positive and negative messages.  The positive message is that the most likely value of the IFR is below 1~\%.  It should however be remarked that the value has been extracted from wealthy countries with state-of-the-art medical care.  It is quite possible that higher rates will be found in other regions. \\

However, the number of cases that have been reported is a small (typically less than 10~\%) fraction of the infected population and this will make contact tracing more difficult.  Indeed, a higher identification fraction would be necessary for maximum effectiveness of contact tracing (see \cite{ref:Model}).  However, there is no need to assume that contact tracing alone will be used to slow or stop the epidemic, and it will form just one of the approaches to limiting the spread of Covid-19 and further novel viruses.  This is clearly foreseen and effective measures such as wearing face masks and adhering to more social distancing will also surely bring a significant positive effect. \\

Finally, it should be clear that there are still very large uncertainties on the parameters extracted in this analysis. A much greater level of random testing will make the parameter inference much more reliable and is strongly supported!

\section{acknowledgements}
We would like to thank a number of colleagues for informative discussions.  In particular, the following colleagues from the Technical University, Munich significantly helped in our evaluations: Elisa Resconi, Stefan Schoenert, Ulrike Protzer and Rudi Zagst.   We would also like to thank Richard Nisius for taking a critical look at an early version of this manuscript and for his valuable comments, and Xiaoyue Li, Olaf Behnke and Michele Giannelli for valualable discussions.

\end{document}